\documentclass[aps,prl,twocolumn,superscriptaddress]{revtex4-1}

\usepackage{graphicx}
\usepackage{subfigure}
\usepackage{amsmath}
\usepackage[colorlinks=true,linkcolor=blue,citecolor=blue, urlcolor=blue]{hyperref}

\newcommand{\dd}{\mathrm{d}}

\newcommand{\ket}[1]{|#1\rangle}

\newcommand{\mbf}[1]{\mathbf{#1}}
\newcommand{\mcal}[1]{\mathcal{#1}}
\newcommand{\mr}[1]{\mathrm{#1}}

\newcounter{fnnumber}

\begin{document}

\title{Exciton Condensation in Landau Levels of Quantum Spin Hall Insulators}

\author{Hong-Mao Peng}
\thanks{These authors contributed equally.}
\affiliation{Kavli Institute for Theoretical Sciences and CAS Center for Excellence in Topological Quantum Computation, University of Chinese Academy of Sciences, Beijing 100190, China}

\author{Zhan Wang}
\thanks{These authors contributed equally.}
\affiliation{Kavli Institute for Theoretical Sciences and CAS Center for Excellence in Topological Quantum Computation, University of Chinese Academy of Sciences, Beijing 100190, China}

\author{Long Zhang}
\email{longzhang@ucas.ac.cn}
\affiliation{Kavli Institute for Theoretical Sciences and CAS Center for Excellence in Topological Quantum Computation, University of Chinese Academy of Sciences, Beijing 100190, China}
\affiliation{Hefei National Laboratory, Hefei 230088, China}

\date{\today}

\begin{abstract}
We theoretically study the quantum spin Hall insulator (QSHI) in a perpendicular magnetic field. In the noninteracting case, the QSHI with space inversion and/or uniaxial spin rotation symmetry undergoes a topological transition into a normal insulator phase at a critical magnetic field $B_{\mr{c}}$. The exciton condensation in the lowest Landau levels is triggered by Coulomb interactions in the vicinity of $B_{\mr{c}}$ at low temperature and spontaneously breaks the inversion and the spin rotation symmetries. We propose that the electron spin resonance spectroscopy with the ac magnetic field also aligned in the perpendicular direction can directly probe the exciton condensation order. Our results should apply to QSHIs such as the InAs/GaSb quantum wells and monolayer transition metal dichalcogenides.
\end{abstract}
\maketitle

\emph{Introduction.}--- The discovery of quantum spin Hall (QSH) effect has opened up a new era of exploring topological materials in the past two decades \cite{Hasan2010, Qi2011}. The QSH state can be constructed by stacking two copies of quantum anomalous Hall (QAH) states, which form a Kramers doublet of the time-reversal symmetry (TRS) with opposite quantized Hall conductance $\sigma_{\mr{H}}=\pm e^{2}/h$ \cite{Kane2005, Bernevig2006}. The QSH insulator (QSHI) has a helical edge state, in which electrons with opposite spin polarization propagate in opposite directions. The single-particle backscattering process is forbidden by the TRS \cite{Kane2005, Wu2006, Xu2006}, thus the edge states are dissipationless and exhibit quantized conductance and nonlocal transport in mesoscopic devices \cite{Konig2007, Roth2009}. The QSH effect has been observed in two-dimensional (2D) semiconductors with an inverted band structure, e.g., the HgTe/CdTe quantum well (QW) \cite{Bernevig2006a, Konig2007}, the InAs/GaSb QW \cite{Liu2008, Knez2011}, the monolayer 1T$'$-WTe$_{2}$ \cite{Qian2014, Tang2017a, Fei2017, Wu2018a}, and the monolayer TaIrTe$_{4}$ \cite{Liu2017c, Guo2020c, Tang2024}.

The excited electron-hole pairs in semiconductors can form bound states called excitons due to the Coulomb interaction. If the binding energy overcomes the band gap, excitons can proliferate and condense at low temperature, forming an exciton insulator (EI) \cite{Blatt1962, Keldysh1965, Jerome1967}. The EI state is well-established in double-layer semiconductors fabricated with GaAs \cite{Eisenstein2004, Eisenstein2014} and bilayer graphene \cite{Liu2017b, Li2017f} in the quantum Hall regime. The interlayer exciton condensation spontaneously breaks the conservation of interlayer electron number difference, and leads to coherent interlayer tunneling \cite{Spielman2000, Spielman2001}, quantized Hall drag resistance \cite{Kellogg2002, Kellogg2003, Liu2017b, Li2017f} and dissipationless counterflow currents \cite{Kellogg2004, Tutuc2004, Li2017f}, which are unambiguous evidence of exciton condensation. The exciton condensation in double-layer fractional quantum Hall states was also proposed recently \cite{Zhang2023k}.

The interplay of band topology and electron interactions can lead to novel quantum states of matter. The possible EI states in the InAs/GaSb QWs \cite{Cheng1995, Du2017a, Wu2019b, Wu2019c, Wang2023} and transition-metal dichalcogenides \cite{Kogar2017, Wang2019, Ma2021, Jia2022, Sun2022a} have been intensively studied recently. In double-layer semiconductors with type-II band alignment, the interplay of chiral interlayer hopping triggered by spin-orbit coupling and exciton condensation driven by Coulomb interaction leads to a rich quantum phase diagram, including topological QSHI and QAHI, normal insulator (NI), and various EI states breaking the TRS and/or certain lattice symmetries \cite{Naveh1996, Pikulin2014, Pikulin2016, Kharitonov2016, Xue2018, Blason2020, Zhu2019b, Amaricci2023}. However, there has been only indirect evidence of EI states in existing experiments reporting the electric transport anomalies \cite{Jia2022, Sun2022a, Wu2019b, Wu2019c, Wang2023} and the presumable spectroscopic exciton gap \cite{Kogar2017, Cheng1995, Du2017a}. More direct probe of the exciton condensation order is still desirable.

\begin{figure}[tb]
\centering
\includegraphics[width=\columnwidth]{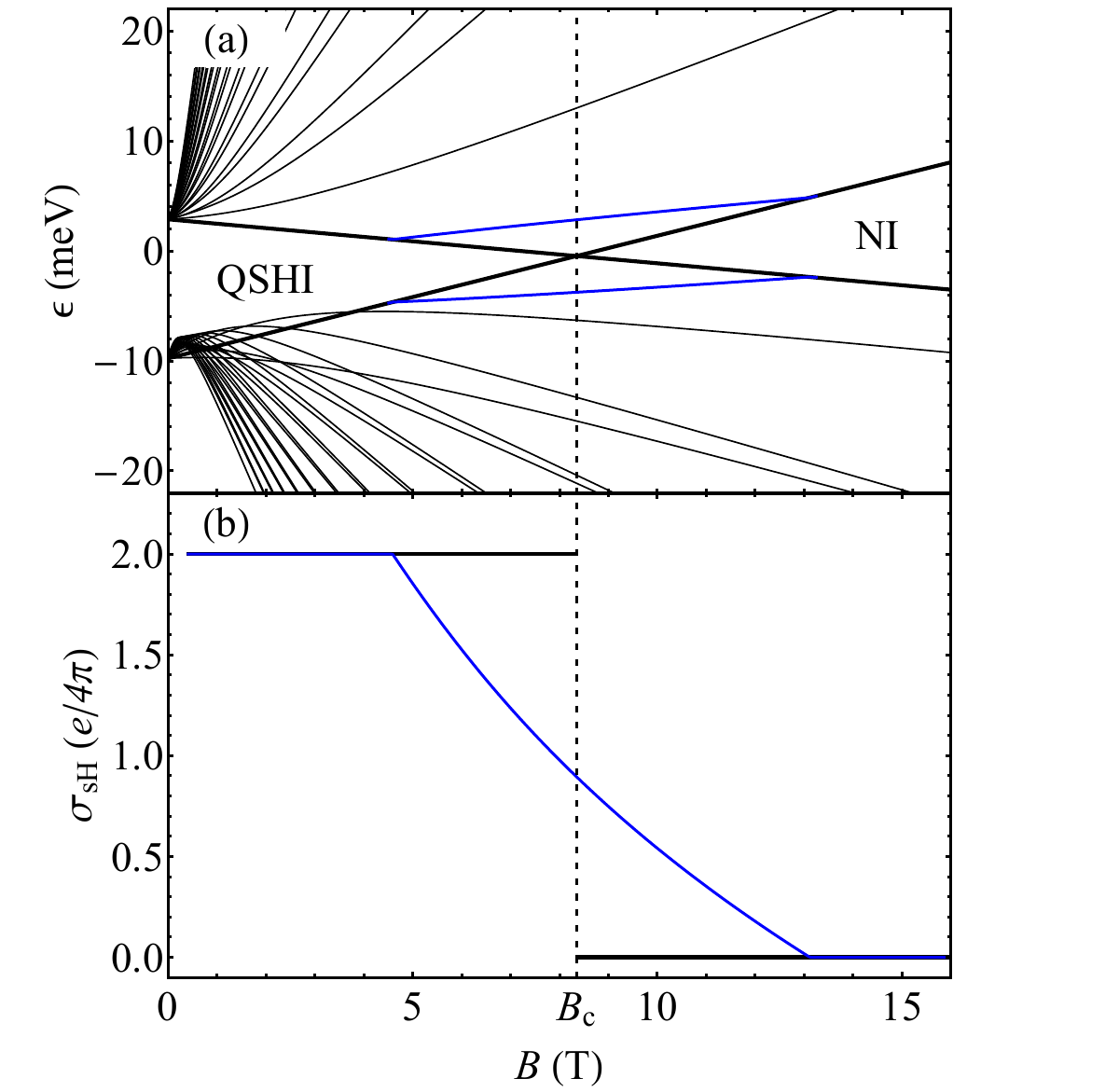}
\caption{(a) The evolution of LLs of the QSHI in a perpendicular magnetic field (black lines). The two lowest LLs (thick black lines) intersect at a critical field $B_{\mr{c}}$, marking a topological transition from the QSHI to the NI state. The blue lines depict the quasiparticle energy levels with exciton condensation in the vicinity of $B_{\mr{c}}$. (b) The spin Hall conductance $\sigma_{\mr{sH}}$ at zero temperature is quantized in the noninteracting case and jumps from $e/(2\pi)$ to zero at $B_{\mr{c}}$ (black lines), but it is not quantized in the EI phase (blue line).}
\label{fig:phase}
\end{figure}

In this work, we focus on the 2D QSHI subject to a perpendicular magnetic field. The QSHI is described by the Bernevig-Hughes-Zhang (BHZ) model \cite{Bernevig2006a}. Even though the TRS is explicitly broken by the magnetic field, the BHZ model has extra space inversion and uniaxial spin rotation symmetries, thus the spin Chern number \cite{Sheng2006} is a well-defined topological invariant, $C_{\mr{s}}=(C_{+}-C_{-})/2$, in which $C_{\alpha}$ ($\alpha=\pm$) are the Chern numbers in different spin sectors. The spin Hall conductance at zero temperature is proportional to the spin Chern number and is quantized as $\sigma_{\mr{sH}}=C_{\mr{s}}\cdot e/(2\pi)$. The QSH state in the BHZ model has $C_{\mr{s}}=1$, and is robust in a perpendicular magnetic field up to a critical field strength $B_{\mr{c}}$, where the two lowest Landau levels (LLs) intersect \cite{Scharf2012, Chen2012d} [see Fig. \ref{fig:phase} (a)]. The LL crossing induces a topological quantum phase transition into a NI phase with $C_{\mr{s}}=0$ above $B_{\mr{c}}$.

The quantum phase diagram is further enriched by the Coulomb interaction. In the vicinity of $B_{\mr{c}}$, the energy gap of electron-hole excitations is vanishingly small, thus it can be easily overcome by the exciton binding energy. Therefore, the exciton condensation takes place and an EI phase intervenes between the QSHI and the NI phases at low temperature \cite{Pikulin2016, Kharitonov2016}. The exciton condensation spontaneously breaks the inversion and the spin rotation symmetries, thereby the spin Chern number is not well-defined any more, and the spin Hall conductance $\sigma_{\mr{sH}}$ is not quantized in the EI phase. Instead, $\sigma_{\mr{sH}}$ continuously decreases from $e/(2\pi)$ to zero [see Fig. \ref{fig:phase} (b)]. We propose that the exciton condensation order can be directly detected with the electron spin resonance (ESR) spectroscopy with the ac magnetic field also aligned in the perpendicular direction. Our results should apply to QSHIs such as the InAs/GaSb quantum wells and monolayer transition metal dichalcogenides.

\emph{Topological transition in BHZ model.}--- The QSHI is described by the four-band BHZ model \cite{Bernevig2006a},
\begin{equation}
H_{0}=\sum_{\mbf{k}}\Psi_{\mbf{k}}^{\dagger}H_{0}(\mbf{k})\Psi_{\mbf{k}},
\end{equation}
in which $\Psi_{\mbf{k}}=(c_{\mbf{k}+},d_{\mbf{k}+},c_{\mbf{k}-},d_{\mbf{k}-})^{\mr{T}}$ denotes the electron operators in the conduction band ($c_{\mbf{k}\alpha}$) and the valence band ($d_{\mbf{k}\alpha}$) with spin index $\alpha=\pm$ \footnote{To be precise, neither the spin nor the total angular momentum is conserved in the semiconductor QWs due to the spin-orbit coupling and the lattice anisotropy \cite{Bernevig2006a}, and the index $\alpha=\pm$ is only a label of the Kramers doublet in each band. Nonetheless, we use the term ``spin'' for convenience.}. In this basis, the Hamiltonian is block-diagonal,
\begin{equation}
H_{0}(\mbf{k})=
\begin{pmatrix}
h(\mbf{k}) & 0 \\
0 & h^{*}(-\mbf{k})
\end{pmatrix},
\end{equation}
and
\begin{equation}
h(\mbf{k})=
\begin{pmatrix}
\frac{\hbar^{2}\mbf{k}^{2}}{2m_{\mr{e}}^{*}}-\mu_{\mr{e}} & w(k_{x}-ik_{y}) \\
w(k_{x}+ik_{y}) & -\frac{\hbar^{2}\mbf{k}^{2}}{2m_{\mr{h}}^{*}}+\mu_{\mr{h}}
\end{pmatrix},
\end{equation}
in which $m_{\mr{e}}^{*}$ and $m_{\mr{h}}^{*}$ are the effective masses of the conduction and the valence bands, respectively. The chiral form of the interband hopping is dictated by the space inversion and the space rotation symmetries \cite{Bernevig2006a}. The system is in the QSH state for $\mu_{\mr{e}}+\mu_{\mr{h}}>0$. If the interband hopping is turned off, the system at the charge neutral point is a semimetal with an equal density of electrons and holes, and the density $n_{0}$ is given by $\mu_{\mr{e}}+\mu_{\mr{h}}=\pi \hbar^{2}n_{0}(1/m_{\mr{e}}^{*}+1/m_{\mr{h}}^{*})$.

The TRS operator $\mcal{T}=-i\sigma_{2}\mcal{K}$ satisfies $\mcal{T}H_{0}(\mbf{k})\mcal{T}^{-1}=H_{0}(-\mbf{k})$, in which $\sigma_{2}$ is the Pauli matrix acting on the spin indices, and $\mcal{K}$ is the complex conjugation. Besides the TRS, the BHZ model has a space inversion symmetry and a uniaxial spin rotation symmetry. The space inversion operator $\mcal{I}=\tau_{3}$ satisfies $\mcal{I}H_{0}(\mbf{k})\mcal{I}^{-1}=H_{0}(-\mbf{k})$, in which $\tau_{3}$ is the Pauli matrix acting on the band indices, because the two bands are derived from $s$- and $p$-like orbitals, respectively \cite{Bernevig2006a}. The uniaxial spin rotation is generated by $J_{z}=\sigma_{3}/2$, which satisfies $[J_{z},H_{0}(\mbf{k})]=0$. In realistic materials, the inversion and the spin rotation symmetries can be explicitly broken by extra asymmetry terms in the Hamiltonian, but these terms are usually much smaller than other relevant energy scales such as the hybridization gap in QSHIs \cite{Liu2008, Liu2013topological}, thus such weak asymmetry can be safely neglected at the leading order. Their effect will be discussed later.

For concreteness, the band parameters of the BHZ model in the following calculations are taken according to experiments in the InAs/GaSb QWs \cite{Mu2016, Han2019a}, $m_{\mr{e}}^{*}=0.04~m_{0}$, $m_{\mr{h}}^{*}=0.136~m_{0}$, $w=0.8~\mr{eV}\cdot\text{\AA}$, and $n_{0}=1.63\times 10^{15}~\mr{m}^{-2}$, in which $m_{0}$ is the free electron mass. These band parameters are highly tunable by changing the QW structure \cite{Liu2008, Liu2013topological, Du2017} and applying gate voltages. The main results in this work are valid as long as the QW is in the inverted regime, and should also apply to other QSHIs with (approximate) inversion and/or uniaxial spin rotation symmetries.

The perpendicular magnetic field enters the Hamiltonian via the Peierls substitution $\mbf{k}\rightarrow \mbf{k}-e\mbf{A}/\hbar$ and the Zeeman coupling $H_{\mr{Z}}=-\frac{1}{2}\mu_{\mr{B}}B\cdot\mr{diag}(g_{\mr{e}},g_{\mr{h}},-g_{\mr{e}},-g_{\mr{h}})$, in which $\mu_{\mr{B}}$ is the Bohr magneton. The $g$-factors $g_{\mr{e}}=11.5$ and $g_{\mr{h}}=0.9$ are also taken from experiments \cite{Mu2016}. In the Landau gauge, the vector potential $\mbf{A}=(0,Bx)$, thus the momentum $k_{y}$ is conserved. Moreover, the Hamiltonian has a magnetic translation symmetry generated by $\pi_{x}=k_{x}-eBy/\hbar$, which shifts $k_{y}$ by a constant, $e^{-i\lambda\pi_{x}}k_{y}e^{i\lambda\pi_{x}}=k_{y}-eB\lambda/\hbar$. We introduce the ladder operators of LLs in the $k_{y}=k$ subspace,
\begin{equation}
a_{k}^{\pm}=\frac{1}{\sqrt{2eB/\hbar}}\big(k_{x}\mp i(k-eBx/\hbar)\big),
\end{equation}
and denote the $n$-th LL with spin $\alpha$ in the conduction band and the valence band by $\ket{\mr{E}_{\alpha},n}$ and $\ket{\mr{H}_{\alpha},n}$, respectively, then the interband hopping term hybridizes $\ket{\mr{E}_{+},n}$ with $\ket{\mr{H}_{+},n-1}$ ($n\ge 1$) in the spin-up sector, and $\ket{\mr{E}_{-},n-1}$ with $\ket{\mr{H}_{-},n}$ ($n\ge 1$) in the spin-down sector. The hybridized LL energies are given by
\begin{gather}
\epsilon_{n}^{+}=\frac{1}{2}(\xi_{n}^{+}+\zeta_{n-1}^{+})\pm\frac{1}{2}\sqrt{(\xi_{n}^{+}-\zeta_{n-1}^{+})^{2}+8neBw^{2}/\hbar}, \\
\epsilon_{n}^{-}=\frac{1}{2}(\xi_{n-1}^{-}+\zeta_{n}^{-})\pm\frac{1}{2}\sqrt{(\xi_{n-1}^{-}-\zeta_{n}^{-})^{2}+8neBw^{2}/\hbar},
\end{gather}
in which $\xi_{n}^{\pm}=\frac{\hbar eB}{m_{\mr{e}}^{*}}(n+1/2)-\mu_{\mr{e}}\mp\frac{1}{2}g_{\mr{e}}\mu_{\mr{B}}B$ and $\zeta_{n}^{\pm}=-\frac{\hbar eB}{m_{\mr{h}}^{*}}(n+1/2)+\mu_{\mr{h}}\mp\frac{1}{2}g_{\mr{h}}\mu_{\mr{B}}B$ are the LL energies in the conduction band and the valence band without hybridization, respectively. The hybridization pushes these LLs away from the Fermi energy at the charge neutral point, which is shown in Fig. \ref{fig:phase} (a). However, the two lowest LLs $\ket{\mr{E}_{+},0}$ and $\ket{\mr{H}_{-},0}$ are not hybridized, $\epsilon_{0}^{+}=\xi_{0}^{+}$ and $\epsilon_{0}^{-}=\zeta_{0}^{-}$. They intersect at a critical magnetic field $B_{\mr{c}}$ \cite{Chen2012d, Scharf2012},
\begin{equation}
B_{\mr{c}}=\frac{\mu_{\mr{e}}+\mu_{\mr{h}}}{\frac{\hbar e}{2m_{\mr{e}}^{*}}-\frac{1}{2}g_{\mr{e}}\mu_{\mr{B}}+\frac{\hbar e}{2m_{\mr{h}}^{*}}-\frac{1}{2}g_{\mr{h}}\mu_{\mr{B}}}.
\end{equation}
This level crossing cannot be avoided, because these two lowest LLs carry opposite quantum numbers of the inversion and the spin rotation symmetries, thus they cannot hybridize with each other without breaking both symmetries.

At the charge neutral point, the LL crossing at $B_{\mr{c}}$ closes the eletron-hole excitation gap, and induces a topological quantum phase transition from the QSHI to the NI state. Although the TRS is explicitly broken, the inversion and the spin rotation symmetries are preserved, thus the topological transition can be characterized by the spin Chern number. Intuitively, the QSHI at zero magnetic field has $C_{\mr{s}}=1$, which persists up to the gap closing at $B_{\mr{c}}$ due to its topological robustness. Above $B_{\mr{c}}$, all LLs from the valence band are filled while all LLs from the conduction band are empty, thus the system is a NI with $C_{\mr{s}}=0$. The spin Chern number can be extracted from the spin Hall conductance at zero temperature calculated with the Kubo formula \cite{Qi2006, Zhang2014d},
\begin{equation}
\sigma_{\mr{sH}}=-\frac{e \hbar^{2}}{2\Omega}\sum_{mn}\mr{Im}\frac{\langle m|\sigma_{3}v_{x}|n\rangle\langle n|v_{y}|m\rangle}{(\epsilon_{m}-\epsilon_{n})^{2}}\big(f(\epsilon_{m})-f(\epsilon_{n})\big),
\label{eq:kubo}
\end{equation}
in which $v_{\alpha}=-(i/\hbar)[r_{\alpha},H]$ ($r_{\alpha}=x,y$) are the electron velocity operators, and $f(\epsilon_{m})$ is the Fermi distribution function of the eigenstate $\ket{m}$ with energy $\epsilon_{m}$. $\sigma_{\mr{sH}}$ is quantized at zero temperature and jumps from $e/(2\pi)$ to zero at $B_{\mr{c}}$ as shown in Fig. \ref{fig:phase} (b), which unambiguously indicates the topological transition from the QSHI to the NI phase.

\begin{figure*}[tb]
\centering
\includegraphics[width=\textwidth]{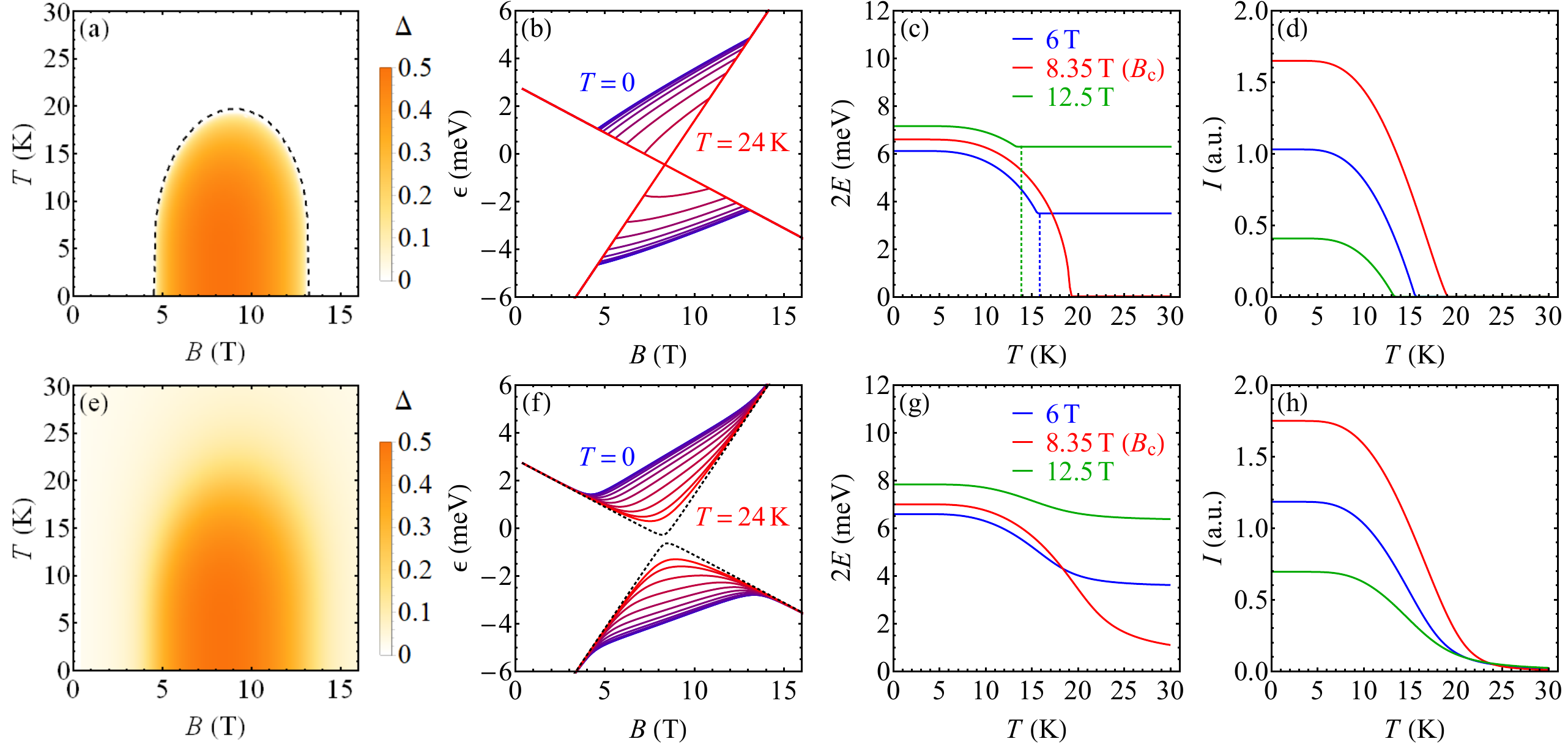}
\caption{(a) Density plot of the exciton condensation order parameter $\Delta$. The dashed line depicts the critical temperature. (b) The hybridized energy levels with exciton condensation at temperatures from $T=0$ (blue lines) to $T=24~\mr{K}$ (red lines). Parameters in the Coulomb potential: $d=10~\mr{nm}$ and $\epsilon_{\mr{r}}=15$. (c) Electron-hole excitation gap $2E$, which equals the ESR resonance frequency in the EI phase, and (d) ESR resonant absorption rate $I$ in arbitrary unit. The critical temperature of exciton condensation is marked by the dashed lines in (c). (e--h) Similar results obtained with a weak inversion asymmetry term $\lambda=0.2~\mr{meV}$. The EI phase boundary is smeared due to the explicit symmetry breaking. The dashed lines in (f) are the hybridized LLs without exciton condensation.}
\label{fig:sol}
\end{figure*}

\emph{Exciton condensation in LLs.}--- The vanishingly small electron-hole excitation gap in the vicinity of $B_{\mr{c}}$ is susceptible to exciton condensation \cite{Pikulin2016, Kharitonov2016}. We may focus on the Coulomb interaction projected into the two lowest LLs, because all other LLs are pushed far away from the Fermi energy,
\begin{equation}
H_{\mr{I}}=\sum_{kk'qq'}h_{k'q'qk}c_{k'}^{\dagger}d_{q'}^{\dagger}d_{q}c_{k},
\end{equation}
in which $c_{k}$ and $d_{k}$ are the electron operators in the two lowest LLs $\ket{\mr{E}_{+},0}$ and $\ket{\mr{H}_{-},0}$ labeled by the momentum $k_{y}=k$, respectively. The interaction coefficients are given by
\begin{equation}
h_{k'q'qk}=\int\dd ^{2}\mbf{r}\dd ^{2}\mbf{r}'\,U(\mbf{r}-\mbf{r}')\psi_{k'}(\mbf{r})\psi_{q'}(\mbf{r}')\psi_{q}^{*}(\mbf{r}')\psi_{k}^{*}(\mbf{r}).
\end{equation}
Here, $\psi_{k}(\mbf{r})$ is the real-space wavefunction in the lowest LL, and $U(\mbf{r})=\gamma/\sqrt{\mbf{r}^{2}+d^{2}}$ is the interlayer Coulomb potential, in which $d$ is the interlayer distance, $\gamma=e^{2}/(4\pi\epsilon_{0}\epsilon_{\mr{r}})$, and $\epsilon_{0}\epsilon_{\mr{r}}$ is the dielectric constant \footnote{See the Supplemental Materials for details on the Coulomb interaction, discussions on more generic symmetry breaking orders, an extended six-LL model, the spin Hall conductance, the effect of inplane magnetic field, and ESR absorption rates versus photon energy, which include Refs. \cite{Qi2006, Zhang2014d, Yang1997}.}\setcounter{fnnumber}{\thefootnote}. The coefficients satisfy $h_{k'+\kappa,q'+\kappa,q+\kappa,k+\kappa}=h_{k'q'qk}$ due to the magnetic translation symmetry and $h_{k'q'qk}=h_{q'k'kq}$ due to the parity symmetry $U(\mbf{r})=U(-\mbf{r})$. The two-LL approximation is justified by an extended six-LL model involving possible competing orders \footnotemark[\thefnnumber].

The possible exciton condensation in the two lowest LLs is captured by the order parameter $\langle c_{k}^{\dagger}d_{q}\rangle=\Delta_{k}\delta_{kq}$, which is assumed to preserve the momentum $k$ in $y$ direction \footnotemark[\thefnnumber]. However, both the inversion and the spin rotation symmetries are spontaneously broken by the exciton condensation, while the composite symmetry $\mcal{I}e^{i\pi J_{z}}$ is preserved; Moreover, if $\Delta_{k}$ depends on $k$, the magnetic translation symmetry is also broken, resulting into a nonuniform exciton condensate. The mean-field Hamiltonian is given by $H_{\mr{MF}}=H_{\mr{LL}}+H_{\mr{I,MF}}$,
\begin{gather}
H_{\mr{LL}} = \sum_{k}(\epsilon_{0}^{+}c_{k}^{\dagger}c_{k}+\epsilon_{0}^{-}d_{k}^{\dagger}d_{k}), \\
H_{\mr{I,MF}} = -\sum_{k}(A_{k}^{*}c_{k}^{\dagger}d_{k}+A_{k}d_{k}^{\dagger}c_{k}),
\end{gather}
in which $A_{k}=\sum_{k'}h_{k'kk'k}\Delta_{k'}$. To solve the mean-field Hamiltonian, we first note that the charge neutrality is maintained by setting the Fermi energy $\mu=(\epsilon_{0}^{+}+\epsilon_{0}^{-})/2$ at any temperature. The coupled self-consistency equations are given by
\begin{equation}
\Delta_{k}=\frac{A_{k}}{2E_{k}}\big(f(-E_{k})-f(E_{k})\big),\quad \forall k,
\label{eq:sce}
\end{equation}
in which $E_{k}=\frac{1}{2}\sqrt{(\epsilon_{0}^{+}-\epsilon_{0}^{-})^{2}+4|A_{k}|^{2}}$ is the quasiparticle excitation energy.

The mean-field equations (\ref{eq:sce}) are solved by a constant order parameter $\Delta_{k}=\Delta$, which represents a uniform exciton condensate preserving the magnetic translation symmetry. The order parameter $\Delta$ and the energy levels hybridized by the exciton condensation at different temperatures and magnetic fields are plotted in Fig. \ref{fig:sol} (a) and (b), respectively. The exciton condensation takes place in a dome-like region in the vicinity of $B_{\mr{c}}$ at low temperature, thus an EI phase intervenes between the QSHI and the NI phases. The LL crossing is avoided due to the symmetry breaking, and the electron-hole excitation gap $2E_{k}=2E$ never closes in the EI phase. The spin Chern number is not well-defined in the EI phase due to the symmetry breaking, but the spin Hall conductance $\sigma_{\mr{sH}}$ can nonetheless be calculated with the Kubo formula (\ref{eq:kubo}). As shown in Fig. \ref{fig:phase} (b), $\sigma_{\mr{sH}}$ at zero temperature is not quantized in the EI phase and continuously decreases from $e/(2\pi)$ to zero.

\emph{ESR spectroscopy of exciton condensation.}--- We propose that the ESR is a direct spectroscopic probe of the exciton condensation order. In ESR experiments, an ac magnetic field is applied to induce resonant transitions between electron states, and the electromagnetic absorption rate is measured. We consider an ac magnetic field also aligned in the perpendicular direction and parallel to the static magnetic field (the Voigt configuration) \cite{Oshikawa2002a}. This configuration respects the space inversion and the uniaxial spin rotation symmetries. While the resonant transition between the two lowest LLs is forbidden by these symmetries, such transition is allowed in the EI phase thanks to the spontaneous symmetry breaking induced by the exciton condensation.

The Zeeman coupling of the two lowest LLs with the ac magnetic field is given by $H_{\mr{ac}}=-\frac{1}{2}\mu_{\mr{B}}B_{\mr{ac}}M_{z}\cos(\omega t)$, in which $B_{\mr{ac}}$ and $\omega$ are the amplitude and the frequency of the ac magnetic field, respectively, and
\begin{equation}
M_{z}=\sum_{k}(g_{\mr{e}}c_{k}^{\dagger}c_{k}-g_{\mr{h}}d_{k}^{\dagger}d_{k}).
\end{equation}
The absorption rate \cite{Oshikawa2002a} $I(\omega)=\frac{1}{8}\omega\mu_{\mr{B}}^{2}B_{\mr{ac}}^{2}\mr{Im}\chi_{zz}(\omega)$, and the dynamic magnetic susceptibility $\chi_{zz}(\omega)$ is given by
\begin{equation}
\begin{split}
\chi_{zz}(\omega) &= -i\int_{0}^{\infty}\dd t\,\big\langle[M_{z}(t),M_{z}(0)]\big\rangle e^{i(\omega+i0^{+})t} \\
&= \sum_{mn}\frac{f(\epsilon_{m})-f(\epsilon_{n})}{\omega+i0^{+}+\epsilon_{m}-\epsilon_{n}}|\langle n|M_{z}|m\rangle|^{2},
\end{split}
\end{equation}
in which $\ket{m}$ denotes the eigenstate of the static Hamiltonian with energy $\epsilon_{m}$.

From the mean-field solution in the EI phase, we find
\begin{equation}
I(\omega)\propto \omega|\Delta|^{2}\coth(\beta E/2)\delta(\omega-2E).
\end{equation}
The resonance frequency is the electron-hole excitation gap $2E$, and the absorption rate is proportional to the exciton condensation order parameter $|\Delta|^{2}$, which are plotted in Fig. \ref{fig:sol} (c,d) \footnotemark[\thefnnumber]. Therefore, the emergence of a resonance peak in the vicinity of $B_{\mr{c}}$ at low temperature is a direct evidence of the spontaneous symmetry breaking induced by the exciton condensation.

\emph{Effect of weak inversion asymmetry.}--- In realistic materials, the inversion and the spin rotation symmetries can be explicitly broken either by the intrinsic bulk and structure asymmetry \cite{Liu2008, Liu2013topological} or by applying an inplane magnetic field in type-II QWs \footnotemark[\thefnnumber], thereby the two lowest LLs are hybridized,
\begin{equation}
H_{\mr{asy}}=-\sum_{k}(\lambda^{*}c_{k}^{\dagger}d_{k}+\lambda d_{k}^{\dagger}c_{k}).
\end{equation}
Adding this term to the mean-field Hamiltonian, the self-consistency equations of the exciton condensation are modified as follows,
\begin{equation}
\Delta_{k}=\frac{A_{k}+\lambda}{2E_{k}'}\big(f(-E_{k}')-f(E_{k}')\big),\quad \forall k,
\end{equation}
in which the modified quasiparticle excitation energy $E_{k}'=\frac{1}{2}\sqrt{(\epsilon_{0}^{+}-\epsilon_{0}^{-})^{2}+4|A_{k}+\lambda|^{2}}$. The modified equations are also solved by a constant $\Delta_{k}=\Delta$; However, as shown in Fig. \ref{fig:sol} (e), $\Delta$ is nonzero at any temperature and magnetic field due to the explicit symmetry breaking, thus the EI phase boundary is smeared. Nonetheless, the exciton condensation order is strong only in the vicinity of $B_{\mr{c}}$ at low temperature and is vanishingly small elsewhere, thus the main feature of the ESR spectroscopy is not significantly affected by the weak inversion asymmetry, which is illustrated in Fig. \ref{fig:sol} (g,h) \footnotemark[\thefnnumber].

\emph{Conclusion.}--- In summary, we theoretically studied the 2D QSHI with the space inversion and the uniaxial spin rotation symmetries in a perpendicular magnetic field. In the noninteracting case, the system undergoes a topological transition from the QSHI to the NI state at a critical magnetic field $B_{\mr{c}}$. The Coulomb interaction triggers the exciton condensation in the vicinity of $B_{\mr{c}}$ at low temperature and spontaneously breaks the inversion and the spin rotation symmetries. We propose that the ESR spectroscopy with the ac magnetic field also aligned in the perpendicular direction (the Voigt configuration) can directly probe the exciton condensation order. Our results should apply to QSHIs with (approximate) inversion and/or uniaxial spin rotation symmetries such as the InAs/GaSb QWs and the monolayer transition metal chalcogenides.

\begin{acknowledgments}
H.-M.P. thanks Jin-Tao Jin for helpful discussions. L.Z. is grateful to Rui-Rui Du, Zhongdong Han and Tingxin Li for related stimulating collaborations, and to Yuan Li and Jia-Wei Mei for helpful discussions. This work is supported by the National Natural Science Foundation of China (No. 12174387), the Chinese Academy of Sciences (Nos. XDB28000000, YSBR-057, and JZHKYPT-2021-08), the University of Chinese Academy of Sciences, and the Innovative Program for Quantum Science and Technology (No. 2021ZD0302600). L.Z. is also supported by the Xiaomi Foundation.
\end{acknowledgments}

\bibliography{../../BibTex/library,../../BibTex/books}
\end{document}


\title{Supplemental Materials for ``Exciton Condensation in Landau Levels of Quantum Spin Hall Insulators''}

\author{Hong-Mao Peng}
\thanks{These authors contributed equally.}
\affiliation{Kavli Institute for Theoretical Sciences and CAS Center for Excellence in Topological Quantum Computation, University of Chinese Academy of Sciences, Beijing 100190, China}

\author{Zhan Wang}
\thanks{These authors contributed equally.}
\affiliation{Kavli Institute for Theoretical Sciences and CAS Center for Excellence in Topological Quantum Computation, University of Chinese Academy of Sciences, Beijing 100190, China}

\author{Long Zhang}
\email{longzhang@ucas.ac.cn}
\affiliation{Kavli Institute for Theoretical Sciences and CAS Center for Excellence in Topological Quantum Computation, University of Chinese Academy of Sciences, Beijing 100190, China}
\affiliation{Hefei National Laboratory, Hefei 230088, China}

\date{\today}

\begin{abstract}
The Supplemental Materials present calculation details of the Coulomb interaction, general discussions on the symmetry breaking orders, an extended mean-field theory with six lowest Landau levels and various possible competing order parameters, the spin Hall conductance, the effect of inplane magnetic field, and ESR absorption rates versus photon energy.
\end{abstract}

\maketitle

\section{Coulomb interaction}

In a type-II semiconductor quantum well, the interlayer Coulomb interaction is described by
\begin{equation}
H_{\mr{I}}=\sum_{\alpha \alpha'}\int\dd ^{2}\mbf{r}\dd ^{2}\mbf{r}'\,U(\mbf{r}-\mbf{r}')c_{\alpha}^{\dagger}(\mbf{r})d_{\alpha'}^{\dagger}(\mbf{r}')d_{\alpha'}(\mbf{r}')c_{\alpha}(\mbf{r}).
\end{equation}
Here, $U(\mbf{r})=\gamma/\sqrt{\mbf{r}^{2}+d^{2}}$ is the interlayer Coulomb potential, in which $d$ is the interlayer distance, $\gamma=e^{2}/(4\pi\epsilon_{0}\epsilon_{\mr{r}})$, and $\epsilon_{0}\epsilon_{\mr{r}}$ is the dielectric constant. $c_{\alpha}^{\dagger}(\mbf{r})$ and $d_{\alpha}^{\dagger}(\mbf{r})$ are the electron creation operators in the conduction and the valence bands, respectively. In the Landau level (LL) basis,
\begin{equation}
c_{\alpha}^{\dagger}(\mbf{r})=\sum_{n}\sum_{k}\psi_{nk}(\mbf{r})c_{nk\alpha}^{\dagger},\quad d_{\alpha}^{\dagger}(\mbf{r})=\sum_{n}\sum_{k}\psi_{nk}(\mbf{r})d_{nk\alpha}^{\dagger},
\end{equation}
in which $\psi_{nk}(\mbf{r})$ is the wavefunction at the $n$-th LL with momentum $k_{y}=k$, while $c_{nk\alpha}^{\dagger}$ and $d_{nk\alpha}^{\dagger}$ are the electron creation operators in the LLs $\ket{\mr{E}_{\alpha},n}$ and $\ket{\mr{H}_{\alpha},n}$, respectively. $H_{\mr{I}}$ can be written in the LL basis,
\begin{equation}
H_{\mr{I}}=\sum_{nn'mm'}\sum_{kk'qq'}\sum_{\alpha\alpha'}h_{k'q'qk}^{n'm'mn}c_{n'k'\alpha}^{\dagger}d_{m'q'\alpha'}^{\dagger}d_{mq\alpha'}c_{nk\alpha},
\label{eq:Coulomb}
\end{equation}
in which the coefficient is given by
\begin{equation}
h_{k'q'qk}^{n'm'mn}=\int\dd ^{2}\mbf{r}\dd ^{2}\mbf{r}'\,U(\mbf{r}-\mbf{r}')\psi_{n'k'}(\mbf{r})\psi_{m'q'}(\mbf{r}')\psi_{mq}^{*}(\mbf{r}')\psi_{nk}^{*}(\mbf{r}).
\end{equation}

In high magnetic field, all LLs except the lowest two $\ket{\mr{E}_{+},0}$ and $\ket{\mr{H}_{-},0}$ are pushed far away from the Fermi energy at the charge neutral point, thus we project $H_{\mr{I}}$ into the lowest two LLs,
\begin{equation}
H_{\mr{I}}=\sum_{kk'qq'}h_{k'q'qk}c_{k'}^{\dagger}d_{q'}^{\dagger}d_{q}c_{k},
\end{equation}
in which the following notations are introduced for simplicity: $c_{k}=c_{0k+}$, $d_{k}=d_{0k-}$, and $h_{k'q'qk}=h_{k'q'qk}^{0000}$,
\begin{equation}
h_{k'q'qk}=\frac{1}{2\pi^{2}l_{B}^{2}}\int\dd ^{2}\mbf{r}\dd ^{2}\mbf{r}'\,U(\mbf{r}-\mbf{r}')e^{ik'y}e^{-\frac{(x-k'l_{B}^{2})^{2}}{2l_{B}^{2}}}e^{iq'y'}e^{-\frac{(x'-q'l_{B}^{2})^{2}}{2l_{B}^{2}}}e^{-iqy'}e^{-\frac{(x'-ql_{B}^{2})^{2}}{2l_{B}^{2}}}e^{-iky}e^{-\frac{(x-kl_{B}^{2})^{2}}{2l_{B}^{2}}},
\end{equation}
in which $l_{B}=\sqrt{\hbar/(eB)}$ is the magnetic length. We find that $h_{k'+\kappa,q'+\kappa,q+\kappa,k+\kappa}=h_{k'q'qk}$, which is consistent with the magnetic translation symmetry. In particular, the coefficient $h_{kk'kk'}$ appearing in the mean-field equation of exciton condensation is given by
\begin{equation}
h_{kk'kk'}=\frac{e^{-(k-k')^{2}l_{B}^{2}/2}}{2\pi^{2}l_{B}^{2}}\int\dd ^{2}\mbf{r}\dd ^{2}\mbf{r}'\,U(\mbf{r}-\mbf{r}')e^{i(k-k')(y-y')}e^{-(x^{2}+x'^{2})/l_{B}^{2}}.
\end{equation}

\section{General discussions on symmetry breaking orders}

Let us give a general discussion on the symmetries and possible competing orders in the quantum spin Hall insulator (QSHI) in an out-of-plane magnetic field. In the LL basis, the single-electron states in the conduction ($\mr{E}$) and the valence ($\mr{H}$) bands are labeled by the LL index $n$, the momentum $k_{y}=k$ and the spin $\alpha=\pm$. We denote the electron creation operators of $|\mr{E}_{\alpha},n,k\rangle$ and $|\mr{H}_{\alpha},n,k\rangle$ by $c_{nk\alpha}^{\dagger}$ and $d_{nk\alpha}^{\dagger}$, respectively, then the symmetry transformations of the space inversion, the uniaxial spin rotation by an angle $\theta$, the space translation by a distance $\lambda$ in $y$ direction and the magnetic translation by a distance $\lambda$ in $x$ direction act as follows:
\begin{enumerate}
\item Space inversion:
\begin{equation}
\mcal{I} c_{nk\alpha}^{\dagger} \mcal{I}^{-1}=(-1)^{n} c_{n,-k,\alpha}^{\dagger},\quad \mcal{I} d_{nk\alpha}^{\dagger} \mcal{I}^{-1}=(-1)^{n+1} d_{n,-k,\alpha}^{\dagger};
\end{equation}

\item Spin rotation:
\begin{equation}
U_{z}(\theta) c_{nk\alpha}^{\dagger} U_{z}(\theta)^{-1}=e^{i\alpha\theta/2} c_{nk\alpha}^{\dagger},\quad U_{z}(\theta) d_{nk\alpha}^{\dagger} U_{z}(\theta)^{-1}=e^{i\alpha\theta/2} d_{nk\alpha}^{\dagger};
\end{equation}

\item Translation:
\begin{equation}
T_{y}(\lambda) c_{nk\alpha}^{\dagger} T_{y}(\lambda)^{-1}=e^{ik\lambda}c_{nk\alpha}^{\dagger},\quad T_{y}(\lambda) d_{nk\alpha}^{\dagger} T_{y}(\lambda)^{-1}=e^{ik\lambda}d_{nk\alpha}^{\dagger};
\end{equation}

\item Magnetic translation:
\begin{equation}
T_{x}(\lambda) c_{nk\alpha}^{\dagger} T_{x}(\lambda)^{-1}=c_{nq\alpha}^{\dagger},\quad T_{x}(\lambda) d_{nk\alpha}^{\dagger} T_{x}(\lambda)^{-1}=d_{nq\alpha}^{\dagger},
\end{equation}
in which $q=k-eB\lambda/\hbar$.
\end{enumerate}

In the QSHI in an out-of-plane magnetic field, $\langle c_{nk\alpha}^{\dagger} d_{mk'\alpha}\rangle =\Delta_{nm\alpha}\delta_{kk'}$ for $n\equiv m+1 \mod 2$ due to the interband hybridization without breaking the above symmetries, in which $\Delta_{nm\alpha}$ does not depend on the momentum $k$. Possible competing orders break one or more of these symmetries. The order parameters can be classified accordingly.
\begin{enumerate}
\item Uniform ordered states without breaking the translation and the magnetic translation symmetries:

\subitem 1.1 States only breaking the space inversion symmetry: $\langle c_{nk\alpha}^{\dagger} d_{mk'\alpha}\rangle =\Delta_{nm\alpha}\delta_{kk'}$ for $n\equiv m \mod 2$, $\langle c_{nk\alpha}^{\dagger} c_{mk'\alpha}\rangle = C_{nm\alpha}\delta_{kk'}$ for $n\equiv m+1 \mod 2$, and $\langle d_{nk\alpha}^{\dagger} d_{mk'\alpha}\rangle =D_{nm\alpha}\delta_{kk'}$ for $n\equiv m+1 \mod 2$.

\subitem 1.2 States only breaking the spin rotation symmetry: $\langle c_{nk\alpha}^{\dagger} d_{m,k',-\alpha}\rangle=\Delta_{nm\alpha} \delta_{kk'}$ for $n\equiv m+1 \mod 2$, $\langle c_{nk\alpha}^{\dagger} c_{m,k',-\alpha}\rangle = C_{nm\alpha}\delta_{kk'}$ for $n\equiv m \mod 2$, and $\langle d_{nk\alpha}^{\dagger} d_{m,k',-\alpha}\rangle= D_{nm\alpha}\delta_{kk'}$ for $n\equiv m \mod 2$. This corresponds to a uniform ferromagnetic state polarized in the $xy$ plane.

\subitem 1.3 States breaking both the space inversion and the spin rotation symmetries: $\langle c_{nk\alpha}^{\dagger} d_{m,k',-\alpha}\rangle= \Delta_{nm\alpha}\delta_{kk'}$ for $n\equiv m \mod 2$, $\langle c_{nk\alpha}^{\dagger} c_{m,k',-\alpha}\rangle= C_{nm\alpha}\delta_{kk'}$ for $n\equiv m+1 \mod 2$, and $\langle d_{nk\alpha}^{\dagger} d_{m,k',-\alpha}\rangle= D_{nm\alpha}\delta_{kk'}$ for $n\equiv m+1 \mod 2$. These order parameters are invariant under the composite symmetry transformation $\mcal{I}U_{z}(\pi)$. This corresponds to the EI state studied in this work, and is also a uniform ferromagnetic state polarized in the $xy$ plane.

\item Spatially nonuniform states breaking the translation and/or the magnetic translation symmetries. We shall restrict our discussions to the simpler cases that the translation symmetry in one direction is not broken. Noting that the translation and the magnetic translation symmetries depend on the gauge choice, we may assume that the translation in $y$ direction and thus the momentum $k$ is preserved. Spatially nonuniform ordered states spontaneously breaking the magnetic translation symmetry in $x$ direction are described by the following order parameters:

\subitem 2.1 States preserving the space inversion and the spin rotation symmetries: $\langle c_{nk\alpha}^{\dagger} d_{mk'\alpha}\rangle= \Delta_{nmk\alpha} \delta_{kk'}$ for $n\equiv m+1 \mod 2$, in which $\Delta_{nmk\alpha}$ depends on the momentum $k$.

\subitem 2.2 States only breaking the space inversion symmetry: $\langle c_{nk\alpha}^{\dagger} d_{mk'\alpha}\rangle =\Delta_{nmk\alpha}\delta_{kk'}$ for $n\equiv m \mod 2$, $\langle c_{nk\alpha}^{\dagger} c_{mk'\alpha}\rangle = C_{nmk\alpha}\delta_{kk'}$ for $n\equiv m+1 \mod 2$, and $\langle d_{nk\alpha}^{\dagger} d_{mk'\alpha}\rangle =D_{nmk\alpha}\delta_{kk'}$ for $n\equiv m+1 \mod 2$.

\subitem 2.3 States only breaking the spin rotation symmetry: $\langle c_{nk\alpha}^{\dagger} d_{m,k',-\alpha}\rangle=\Delta_{nmk\alpha} \delta_{kk'}$ for $n\equiv m+1 \mod 2$, $\langle c_{nk\alpha}^{\dagger} c_{m,k',-\alpha}\rangle = C_{nmk\alpha}\delta_{kk'}$ for $n\equiv m \mod 2$, and $\langle d_{nk\alpha}^{\dagger} d_{m,k',-\alpha}\rangle= D_{nmk\alpha}\delta_{kk'}$ for $n\equiv m \mod 2$. This corresponds to a nonuniform ferromagnetic state polarized in the $xy$ plane.

\subitem 2.4 States breaking both the space inversion and the spin rotation symmetries: $\langle c_{nk\alpha}^{\dagger} d_{m,k',-\alpha}\rangle= \Delta_{nmk\alpha}\delta_{kk'}$ for $n\equiv m \mod 2$, $\langle c_{nk\alpha}^{\dagger} c_{m,k',-\alpha}\rangle= C_{nmk\alpha}\delta_{kk'}$ for $n\equiv m+1 \mod 2$, and $\langle d_{nk\alpha}^{\dagger} d_{m,k',-\alpha}\rangle= D_{nmk\alpha}\delta_{kk'}$ for $n\equiv m+1 \mod 2$. This corresponds to the nonuniform EI state studied by the extended mean-field theory, and is also a nonuniform ferromagnetic state polarized in the $xy$ plane.
\end{enumerate}

These order parameters emerge spontaneously if the Coulomb interaction overcomes the corresponding inter-LL gaps. In a strong magnetic field and particularly near the critical field $B_{\mr{c}}$, the LLs are pushed far away from the Fermi energy except for the two lowest LLs $|\mr{E}_{+},0\rangle$ and $|\mr{H}_{-},0\rangle$. Therefore, the exciton condensation order $\langle c_{0k+}^{\dagger} d_{0k'-}\rangle= \Delta_{00k+}\delta_{kk'}$ is the predominant instability, which breaks both the space inversion and the spin rotation symmetries. It describes the uniform EI state if $\Delta_{00k+}$ does not depend on $k$ and the nonuniform EI state otherwise. This is studied with the extended mean-field theory projected into the two lowest LLs in this work, and we find that the uniform EI state is always the stable solution of the self-consistent equations.

\section{Extended six-Landau-level mean-field theory}

In order to investigate the various possible competing orders discussed in the previous section, we turn to an extended model with six lowest LLs, $\ket{\mr{E}_{+},0}$, $\ket{\mr{H}_{-},0}$, $\ket{\mr{E}_{+},1}$, $\ket{\mr{H}_{+},0}$, $\ket{\mr{E}_{-},0}$ and $\ket{\mr{H}_{-},1}$. The Bernevig-Hughes-Zhang Hamiltonian projected in these LLs is given by
\begin{equation}
H_{0}=\sum_{k}\Phi_{k}^{\dagger}
\begin{pmatrix}
\xi_{0}^{+}	& 				& 					& 					& 					& 					\\
			& \zeta_{0}^{-}	& 					& 					& 					&					\\
			& 				& \xi_{1}^{+} 		& \sqrt{2}w/l_{B}	& 					&				 	\\
			& 				& \sqrt{2}w/l_{B}	& \zeta_{0}^{+}		& 					& 					\\
			& 				& 					& 					& \xi_{0}^{-}		& -\sqrt{2}w/l_{B}	\\
			& 				& 					& 					& -\sqrt{2}w/l_{B}	& \zeta_{1}^{-}
\end{pmatrix}
\Phi_{k}.
\end{equation}
Here, $\Phi_{k}=(c_{0k+},d_{0k-},c_{1k+},d_{0k+},c_{0k-},d_{1k-})^{\mr{T}}$, in which $c_{nk\sigma}$ and $d_{nk\sigma}$ denote the electron operators of $\ket{\mr{E}_{\sigma},n}$ and $\ket{\mr{H}_{\sigma},n}$, respectively.

The possible order parameters in these LLs can be classified according to the broken symmetries:
\begin{itemize}
\item Order parameters breaking both the space inversion symmetry $\mcal{I}$ and the spin rotation symmetry $U_{z}(\theta)$ while preserving the composite symmetry $\mcal{I}U_{z}(\pi)$,
\begin{equation}
\Delta_{1}(k)=\langle c_{0k+}^{\dagger}d_{0k-}\rangle, \quad
\Delta_{2}(k)=\langle d_{0k+}^{\dagger}c_{0k-}\rangle, \quad
\Delta_{3}(k)=\langle c_{1k+}^{\dagger}d_{1k-}\rangle,
\end{equation}
in which $\Delta_{1}(k)$ is exactly the exciton condensation order parameter $\Delta_{k}$ studied in the two-LL model.

\item Order parameters breaking the spin rotation symmetry $U_{z}(\theta)$ while preserving the space inversion $\mcal{I}$,
\begin{equation}
\Delta_{4}(k)=\langle c_{0k+}^{\dagger}d_{1k-}\rangle, \quad
\Delta_{5}(k)=\langle d_{0k-}^{\dagger}c_{1k+}\rangle.
\end{equation}

\item Order parameters breaking the space inversion symmetry $\mcal{I}$ while preserving the spin rotation symmetry $U_{z}(\theta)$,
\begin{equation}
\Delta_{6}(k)=\langle c_{0k+}^{\dagger}d_{0k+}\rangle, \quad
\Delta_{7}(k)=\langle d_{0k-}^{\dagger}c_{0k-}\rangle, \quad
\Delta_{8}(k)=\langle c_{0k+}^{\dagger}c_{1k+}\rangle, \quad
\Delta_{9}(k)=\langle d_{0k-}^{\dagger}d_{1k-}\rangle.
\end{equation}
\end{itemize}
These order parameters are in general $k$-dependent, thus they can describe the possible nonuniform states breaking the magnetic translation symmetry.

The interlayer Coulomb interaction Eq. (\ref{eq:Coulomb}) is projected into the six LLs, and is decoupled by introducing these order parameters. The mean-field Hamiltonian is given by
\begin{equation}
H_{\mr{MF}}=\sum_{k}\Phi_{k}^{\dagger}H_{\mr{MF}}(k)\Phi_{k},
\end{equation}
and
\begin{equation}
H_{\mr{MF}}(k)=
\begin{pmatrix}
\xi_{0}^{+}		& h_{12}(k)		& h_{13}(k)			& h_{14}(k)			& 0					& h_{16}(k)			\\
h_{12}^{*}(k)	& \zeta_{0}^{-}	& h_{23}(k)			& 0					& h_{25}(k)			& h_{26}(k)			\\
h_{13}^{*}(k)	& h_{23}^{*}(k)	& \xi_{1}^{+}		& \sqrt{2}w/l_{B}	& 0					& h_{36}(k)			\\
h_{14}^{*}(k)	& 0				& \sqrt{2}w/l_{B}	& \zeta_{0}^{+}		& h_{45}(k)			& 0					\\
0				& h_{25}^{*}(k)	& 0					& h_{45}^{*}(k)		& \xi_{0}^{-}		&-\sqrt{2}w/l_{B}	\\
h_{16}^{*}(k)	& h_{26}^{*}(k)	& h_{36}^{*}(k)		& 0					&-\sqrt{2}w/l_{B}	& \zeta_{1}^{-}
\end{pmatrix},
\end{equation}
in which the matrix elements are given by
\begin{align}
& h_{12}(k)=\sum_{k'} h^{0000}_{k k' k  k'}\Delta_{1}^{*}(k')	+\sum_{k} h^{0101}_{k k' k  k'}\Delta_{3}^{*}(k'), \\
& h_{13}(k)=\sum_{k'} h^{0110}_{k k  k' k'}\Delta_{9} (k')		+\sum_{k} h^{0101}_{k k' k  k'}\Delta_{9}^{*}(k'), \\
& h_{14}(k)=\sum_{k'} h^{0000}_{k k' k  k'}\Delta_{6}^{*}(k'), \\
& h_{16}(k)=\sum_{k'} h^{0011}_{k k' k  k'}\Delta_{4}^{*}(k')	+\sum_{k} h^{0110}_{k k' k  k'}\Delta_{5}(k'), \\
& h_{23}(k)=\sum_{k'} h^{0110}_{k k' k  k'}\Delta_{4}(k')	+\sum_{k} h^{1100}_{k k' k  k'}\Delta_{5}^{*}(k'), \\
& h_{25}(k)=\sum_{k'} h^{0000}_{k k' k  k'}\Delta_{7}^{*}(k'), \\
& h_{26}(k)=\sum_{k'} h^{0110}_{k k  k' k'}\Delta_{8}(k')		+\sum_{k} h^{1010}_{k k' k  k'}\Delta_{8}^{*}(k'), \\
& h_{36}(k)=\sum_{k'} h^{1111}_{k k' k  k'}\Delta_{3}^{*}(k')	+\sum_{k} h^{1010}_{k k' k  k'}\Delta_{1}^{*}(k'), \\
& h_{45}(k)=\sum_{k'} h^{0000}_{k k' k  k'}\Delta_{2}^{*}(k').
\end{align}

The mean-field Hamiltonian $H_{\mr{MF}}(k)$ is diagonalized with a unitary matrix $U(k)$, $H_{\mr{MF}}(k)=U(k)\Lambda(k)U(k)^{\dagger}$, in which $\Lambda(k)=\mr{diag}(E_{1k},E_{2k},E_{3k},E_{4k},E_{5k},E_{6k})$, and $E_{\alpha k}$ ($1\leq \alpha\leq 6$) are the energy eigenvalues. The self-consistency equations for the order parameters are given by
\begin{gather}
\Delta_{1}(k)= \sum_{\alpha} U_{1\alpha}(k)U_{2\alpha}^{*}(k) n_{\mr{F}}(E_{\alpha k}), \\
\Delta_{2}(k)= \sum_{\alpha} U_{4\alpha}(k)U_{5\alpha}^{*}(k) n_{\mr{F}}(E_{\alpha k}), \\
\Delta_{3}(k)= \sum_{\alpha} U_{3\alpha}(k)U_{6\alpha}^{*}(k) n_{\mr{F}}(E_{\alpha k}), \\
\Delta_{4}(k)= \sum_{\alpha} U_{1\alpha}(k)U_{6\alpha}^{*}(k) n_{\mr{F}}(E_{\alpha k}), \\
\Delta_{5}(k)= \sum_{\alpha} U_{2\alpha}(k)U_{3\alpha}^{*}(k) n_{\mr{F}}(E_{\alpha k}), \\
\Delta_{6}(k)= \sum_{\alpha} U_{1\alpha}(k)U_{4\alpha}^{*}(k) n_{\mr{F}}(E_{\alpha k}), \\
\Delta_{7}(k)= \sum_{\alpha} U_{2\alpha}(k)U_{5\alpha}^{*}(k) n_{\mr{F}}(E_{\alpha k}), \\
\Delta_{8}(k)= \sum_{\alpha} U_{1\alpha}(k)U_{3\alpha}^{*}(k) n_{\mr{F}}(E_{\alpha k}), \\
\Delta_{9}(k)= \sum_{\alpha} U_{2\alpha}(k)U_{6\alpha}^{*}(k) n_{\mr{F}}(E_{\alpha k}),
\end{gather}
in which $n_{\mr{F}}(E_{\alpha k})=1/\big(e^{\beta (E_{\alpha}-\mu)}+1\big)$ is the Fermi-Dirac distribution function. These coupled equations are solved iteratively at different temperatures and magnetic fields, while the chemical potential $\mu$ is adjusted to the charge neutral point during the iteration.

\begin{figure}[!tb]
\centering
\includegraphics[width=\textwidth]{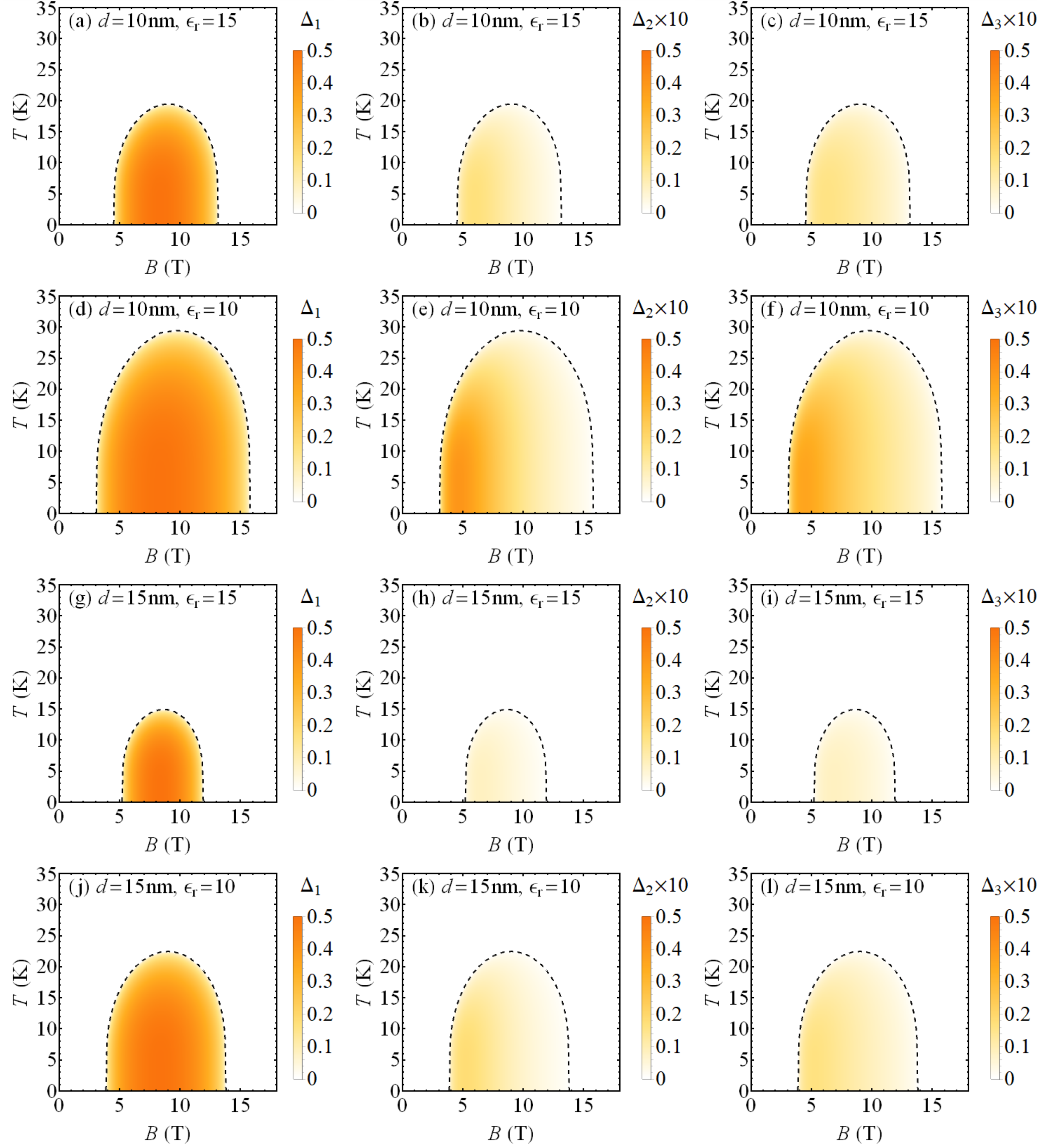}
\caption{Density plots of the order parameters $\Delta_{1,2,3}$ at different temperatures and magnetic fields. Other order parameters $\Delta_{\alpha}$ ($4\leq \alpha\leq 9$) are always zero. Parameters in the Coulomb potential are labeled in each panel. Dashed lines depict the critical temperatures.}
\label{fig:extended}
\end{figure}

For the band parameters of the InAs/GaSb quantum well adopted in the main text, we find that the order parameters are always independent of $k$, which describe a uniform state preserving the magnetic translation symmetry. The results calculated with several sets of Coulomb potential parameters are plotted in Fig. \ref{fig:extended}. The order parameters $\Delta_{\alpha}$ ($4\leq \alpha\leq 9$) are always zero across the whole phase diagram, while $\Delta_{\alpha}$ ($1\leq \alpha\leq 3$) are nonzero in a dome-like region near the critical magnetic field $B_{\mr{c}}$. Therefore, the exciton condensation in the lowest two LLs is the predominant instability in this system. Moreover, we find that $\Delta_{1}$ has almost the same value as the order parameter $\Delta$ in the two-LL model, while $\Delta_{2}$ and $\Delta_{3}$ are much smaller than $\Delta_{1}$. This fully justifies the two-LL approximation we used in the main text.

\section{Spin Hall conductance}	

The spin Hall conductance can be calculated with the Kubo formula \cite{Qi2006, Zhang2014d},
\begin{equation}
\sigma_{\mr{sH}}=-\frac{e \hbar^{2}}{2\Omega}\sum_{mn}\mr{Im}\frac{\langle m|\sigma_{3}v_{x}|n\rangle\langle n|v_{y}|m\rangle}{(\epsilon_{m}-\epsilon_{n})^{2}}\big(f(\epsilon_{m})-f(\epsilon_{n})\big),
\label{eq:kubo}
\end{equation}
in which $v_{\alpha}=-(i/\hbar)[r_{\alpha},H]$ ($r_{\alpha}=x,y$) are the electron velocity operators, and $f(\epsilon_{m})$ is the Fermi distribution function of the eigenstate $\ket{m}$ with energy $\epsilon_{m}$. In the Landau gauge, the velocity operators commute with the momentum $k_{y}$, $[v_{\alpha},k_{y}]=0$. In the subspace with $k_{y}=k$, the velocity operators are given by
\begin{equation}
\hbar v_{x}=
\begin{pmatrix}
\frac{\hbar^2}{\sqrt{2}m_{\mr{e}} l_{B}}(a_{k}^{+}+a_{k}^{-})	& w		& 0	& 0		\\
w	& -\frac{\hbar^2}{\sqrt{2}m_{\mr{h}} l_{B}}(a_{k}^{+}+a_{k}^{-})	& 0	& 0		\\
0	& 0	& \frac{\hbar^2}{\sqrt{2}m_{\mr{e}} l_{B}}(a_{k}^{+}+a_{k}^{-})	& -w	\\
0	& 0	& -w	& -\frac{\hbar^2}{\sqrt{2}m_{\mr{h}} l_{B}}(a_{k}^{+}+a_{k}^{-})
\end{pmatrix},
\end{equation}

\begin{equation}
\hbar v_{y}=
\begin{pmatrix}
\frac{-i\hbar^2}{\sqrt{2}m_{\mr{e}} l_{B}}(a_{k}^{+}-a_{k}^{-})	& iw	& 0	& 0		\\
-iw	& \frac{i\hbar^2}{\sqrt{2}m_{\mr{h}} l_{B}}(a_{k}^{+}-a_{k}^{-})		& 0	& 0		\\
0	& 0	& \frac{-i\hbar^2}{\sqrt{2}m_{\mr{e}} l_{B}}(a_{k}^{+}-a_{k}^{-})		& iw	\\
0	& 0	& -iw	& \frac{i\hbar^2}{\sqrt{2}m_{\mr{h}} l_{B}}(a_{k}^{+}-a_{k}^{-})
\end{pmatrix}.
\end{equation}
Degenerate states in the same LL contribute equally in the Kubo formula (\ref{eq:kubo}), thus the spatial area $\Omega$ is canceled by the LL degeneracy $\Omega\cdot \frac{eB}{2\pi\hbar}$,
\begin{equation}
\sigma_{\mr{sH}}=-\frac{e}{4\pi}\sum_{MN}\mr{Im}\frac{\langle M|\sigma_{3}v_{x}|N\rangle\langle N|v_{y}|M\rangle}{(\epsilon_{M}-\epsilon_{N})^{2}}\big(f(\epsilon_{M})-f(\epsilon_{N})\big),
\end{equation}
in which $M$ and $N$ are the indices of hybridized LLs.

\section{Effect of inplane magnetic field}

In a type-II semiconductor quantum well, the conduction and the valence bands are spatially separated by a distance $d$ in $z$ direction, thus an inplane magnetic field $B_{x}$ couples to the orbital motion of electrons as well. It turns out to break both the space inversion and the uniaxial spin rotation symmetries, and hybridizes the two lowest LLs $\ket{\mr{E}_{+},0}$ and $\ket{\mr{H}_{-},0}$ in the effective Hamiltonian.

In the Landau gauge, the vector potential is given by $\mbf{A}=(0,B_{z}x-B_{x}z)$, in which $z=\pm d/2$ for the conduction and the valence bands, respectively \cite{Yang1997}. With the Peierls substitution $\mbf{k}\rightarrow \mbf{k}-e\mbf{A}/\hbar$, the inplane magnetic field $B_{x}$ introduces an extra term in the orbital motion of electrons,
\begin{equation}
H_{1}'=
\begin{pmatrix}
\frac{(\hbar k_{y}-eB_{z}x)eB_{x} d+(eB_{x} d/2)^2}{2 m_{\mr{e}}}	& 0	& 0 & 0 \\
0	& \frac{(\hbar k_{y}-eB_{z}x)eB_{x} d-(eB_{x} d/2)^2}{2 m_{\mr{h}}}	& 0	& 0 \\
0	& 0	&\frac{(\hbar k_{y}-eB_{z}x)eB_{x} d+(eB_{x} d/2)^2}{2 m_{\mr{e}}}	& 0	\\
0	& 0	& 0	& \frac{(\hbar k_{y}-eB_{z}x)eB_{x} d-(eB_{x} d/2)^2}{2 m_{\mr{h}}}
\end{pmatrix},
\end{equation}
which commutes with the momentum $k_{y}$. In the subspace with $k_{y}=k$,
\begin{equation}
H_{1}'=
\begin{pmatrix}
\frac{i\hbar (a_{k}^{+}-a_{k}^{-})eB_{x} d}{2\sqrt{2} l_{B} m_{\mr{e}}}+\frac{(eB_{x}d/2)^2}{2 m_{\mr{e}}}	& 0	& 0	& 0	\\
0	& \frac{i\hbar (a_{k}^{+}-a_{k}^{-})eB_{x} d}{2\sqrt{2} l_{B} m_{\mr{h}}}-\frac{(eB_{x}d/2)^2}{2 m_{\mr{h}}}& 0	& 0	\\
0	& 0	& \frac{i\hbar (a_{k}^{+}-a_{k}^{-})eB_{x} d}{2\sqrt{2} l_{B} m_{\mr{e}}}+\frac{(eB_{x}d/2)^2}{2 m_{\mr{e}}}& 0	\\
0	& 0	& 0	& \frac{i\hbar (a_{k}^{+}-a_{k}^{-})eB_{x} d}{2\sqrt{2} l_{B} m_{\mr{h}}}-\frac{(eB_{x}d/2)^2}{2 m_{\mr{h}}}
\end{pmatrix},
\end{equation}
thus it can change the LL index by $\pm 1$ without affecting the spin. The Zeeman term of the inplane magnetic field is given by
\begin{equation}
H_{2}'=
\begin{pmatrix}
0	& 0	& -\frac{1}{2}g_{\mr{e}} \mu_{\mr{B}} B_{x}	& 0	\\
0	& 0	& 0	& -\frac{1}{2}g_{\mr{h}} \mu _{\mr{B}} B_{x}\\
-\frac{1}{2}g_{\mr{e}} \mu_{\mr{B}} B_{x}	& 0	& 0	& 0	\\
0	& -\frac{1}{2}g_{\mr{h}} \mu _{\mr{B}} B_{x}& 0	& 0
\end{pmatrix},
\end{equation}
which flips the spin without changing the LL index. Therefore, the two lowest LLs are hybridized via the following second-order processes,
\begin{gather}
\ket{\mr{E}_{+},0}\xlongrightarrow{H_{1}'} \big(\ket{\mr{E}_{+},1},\ket{\mr{H}_{+},0}\big)\xlongrightarrow{H_{2}'} \ket{\mr{H}_{-},0}, \\
\ket{\mr{E}_{+},0}\xlongrightarrow{H_{2}'} \big(\ket{\mr{E}_{-},0},\ket{\mr{H}_{-},1}\big)\xlongrightarrow{H_{1}'} \ket{\mr{H}_{-},0},
\end{gather}
in which the intermediate states are LLs hybridized by the interlayer hopping term, which are away from the Fermi energy.

\begin{figure}[tb]
\centering
\includegraphics[width=0.5\textwidth]{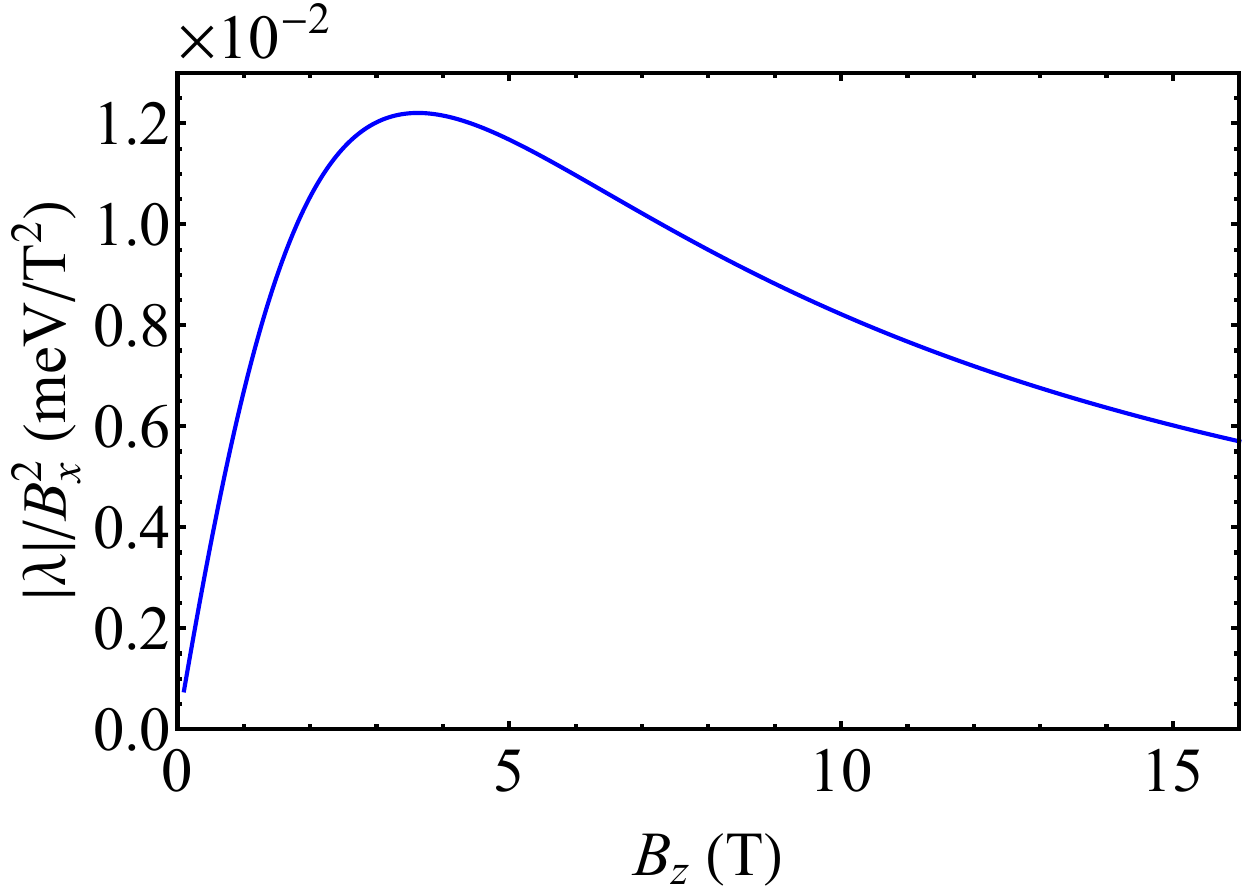}
\caption{The coefficient of the off-diagonal term in the effective Hamiltonian $|\lambda|/B_{x}^{2}$ versus the perpendicular magnetic field $B_{z}$.}
\label{fig:lambda}
\end{figure}

The effective Hamiltonian in the subspace of the two lowest LLs can be derived by the second-order perturbation in $H'=H_{1}'+H_{2}'$,
\begin{equation}
H_{\mr{eff}}=\mbb{P}H_{0}\mbb{P}+\mbb{P}H'\mbb{P}-\mbb{P}H'\mbb{Q}H_{0}^{-1}\mbb{Q}H'\mbb{P},
\end{equation}
in which $\mbb{P}$ is the projection operator into the subspace of $\ket{\mr{E}_{+},0}$ and $\ket{\mr{H}_{-},0}$, and $\mbb{Q}=\mbb{I}-\mbb{P}$ is the projection operator into the subspace of other LLs away from the Fermi energy. This leads to an off-diagonal term in the effective Hamiltonian,
\begin{equation}
H_{\mr{asy}}=-\sum_{k}(\lambda^{*}c_{k}^{\dagger}d_{k}+\lambda d_{k}^{\dagger}c_{k}),
\end{equation}
in which
\begin{equation}
\begin{split}
\lambda &=\frac{i\frac{\hbar e d}{4l_{B}^2  m_{\mr{e}}}w g_{\mr{h}} \mu _{\mr{B}} B_{x}^{2}}
{\big(-\frac{\hbar^2 }{2l_{B}^2  m_{\mr{h}}}+\mu _{\mr{h}}-\frac{1}{2} g_{\mr{h}} \mu _{\mr{B}} B_z\big)\big(\frac{3 \hbar^2 }{2l_{B}^2  m_{\mr{e}}}-\mu_{\mr{e}}-\frac{1}{2} g_{\mr{e}} \mu_{\mr{B}} B_z\big)-\frac{2}{l_{B}^2}w^2} \\
&+\frac{i\frac{\hbar e d}{ 4l_{B}^2  m_{\mr{h}}}w g_{\mr{e}} \mu_{\mr{B}} B_x^2}{\big(-\frac{3\hbar^2 }{2l_{B}^2  m_{\mr{h}}}+\mu_{\mr{h}}+\frac{1}{2} g_{\mr{h}} \mu_{\mr{B}} B_z\big)\big(\frac{ \hbar^2 }{2l_{B}^2  m_{\mr{e}}}-\mu_{\mr{e}}+\frac{1}{2} g_{\mr{e}} \mu_{\mr{B}} B_z\big)-\frac{2}{l_{B_z}^2}w^2}.
\end{split}
\end{equation}
We find $\lambda\propto B_{x}^{2}$ due to the second-order perturbation, and the coefficient depends on the perpendicular magnetic field $B_{z}$, which is plotted in Fig. \ref{fig:lambda}.

\begin{figure}[tb]
\centering
\includegraphics[width=0.8\textwidth]{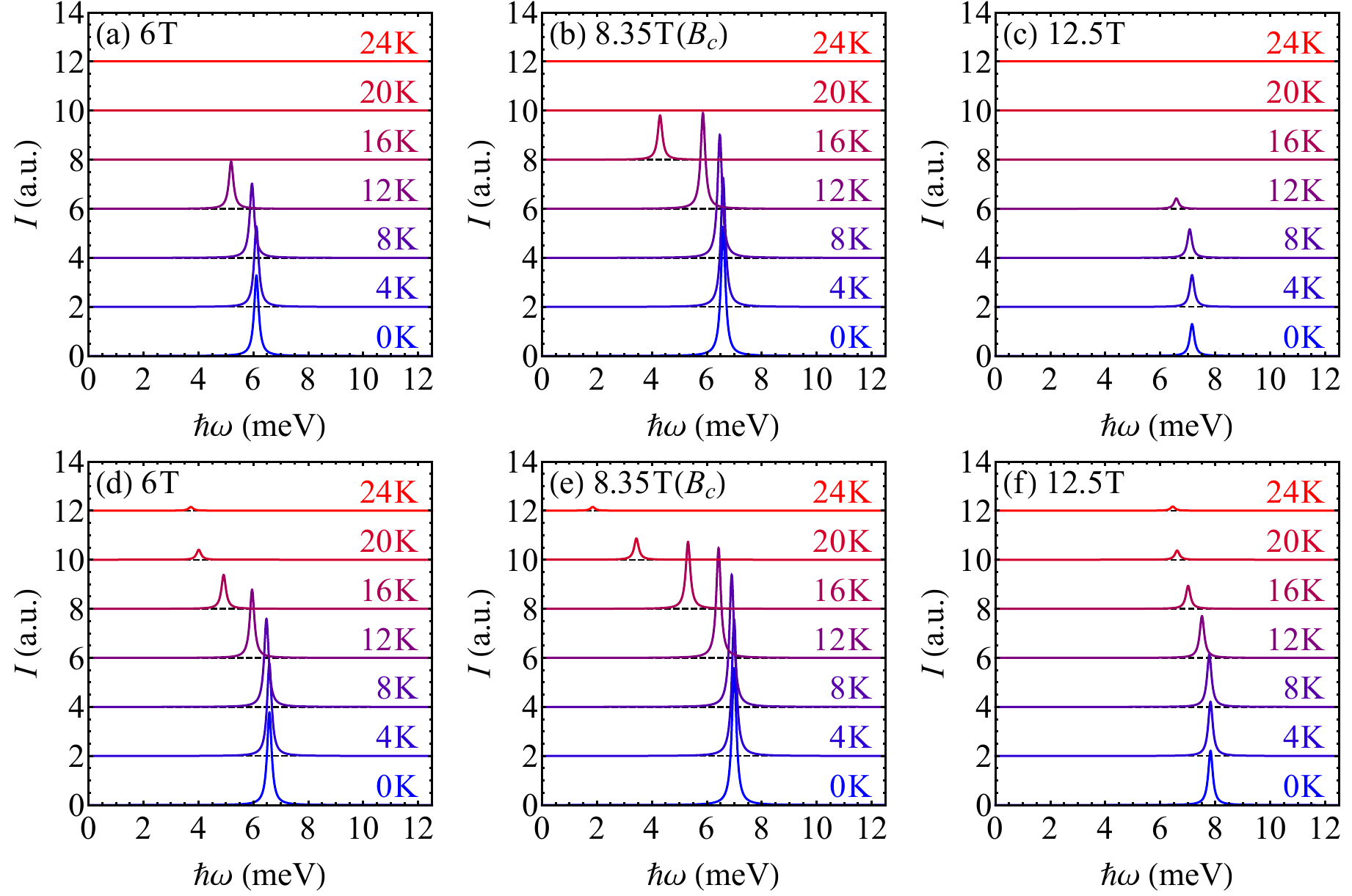}
\caption{The ESR absorption rate $I$ in arbitrary unit (a.u.) versus the photon energy $\hbar\omega$ at different temperatures and magnetic fields (a--c) without inversion asymmetry and (d--f) with a weak inversion asymmetry term $\lambda=0.2~\mr{meV}$. A Lorentz broadening $\eta=0.1~\mr{meV}$ is adopted to illustrate the $\delta$-peaks. The absorption rate curves at different temperatures are vertically shifted for clarity.}
\label{fig:spec}
\end{figure}

\section{ESR absorption rates versus photon energy}

From the mean-field theory, the electron spin resonance (ESR) absorption rate $I(\omega)$ is given by
\begin{equation}
I(\omega)\propto \omega|\Delta|^{2}\coth(\beta E/2)\delta(\omega-2E).
\end{equation}
The resonance frequency is the electron-hole excitation gap $2E$, and the absorption rate is proportional to the exciton condensation order parameter $|\Delta|^{2}$. The ESR absorption rates versus the photon energy at different temperatures and magnetic fields are plotted in Fig. \ref{fig:spec}.

\bibliography{../../BibTex/library,../../BibTex/books}